\documentclass{emulateapj}

\shorttitle{SN 2006gy:  Still Going Strong}
\shortauthors{Smith et al.}
\begin{document}

\title{Late-time observations of SN~2006\lowercase{gy}: Still Going
Strong}

\author{Nathan Smith\altaffilmark{1}, Ryan J.\ Foley\altaffilmark{1},
Joshua S.\ Bloom\altaffilmark{1.2}, Weidong Li\altaffilmark{1}, Alexei
V.\ Filippenko\altaffilmark{1}, \\ Rapha\"el Gavazzi\altaffilmark{3},
Andrea Ghez\altaffilmark{4}, Quinn Konopacky\altaffilmark{4}, Matthew
A.\ Malkan\altaffilmark{4}, \\ Philip J.\ Marshall\altaffilmark{3},
David Pooley\altaffilmark{5}, Tommaso Treu\altaffilmark{2,3,6}, and
Jong-Hak~Woo\altaffilmark{3}}

\altaffiltext{1}{Department of Astronomy, University of California,
Berkeley, CA 94720-3411.}

\altaffiltext{2}{Sloan Research Fellow.}

\altaffiltext{3}{Department of Physics, University of California,
Santa Barbara, CA 93106-9530.}

\altaffiltext{4}{Division of Astronomy and Astrophysics, University of
California, Los Angeles, CA 90095-1547.}

\altaffiltext{5}{Astronomy Department, University of Wisconsin,
Madison, WI 53706.}

\altaffiltext{6}{Packard Fellow.}

\begin{abstract}

Owing to its extremely high luminosity and long duration, supernova
(SN) 2006gy radiated more energy in visual light than any other known
SN.  Two hypotheses to explain its high luminosity at early times ---
that it was powered by shock interaction with circumstellar material
(CSM) as implied by its Type IIn spectrum, or that it was fueled by
radioactive decay from a large mass of $^{56}$Ni synthesized in a
pair-instability SN --- predicted different late-time properties.
Here we present observations of SN~2006gy obtained more than a year
after discovery.  We were unable to detect it at visual wavelengths,
but clear near-infrared (IR) $K\arcmin$ and $H$-band detections show
that it is still at least as luminous as the peak of a normal Type II
SN.  We also present spectra giving an upper limit to the late-time
H$\alpha$ luminosity of $\la$10$^{39}$ erg s$^{-1}$.  Based on the
weak late-time H$\alpha$, X-ray, and radio emission, combined with the
difficulty of explaining the shift to IR wavelengths, we can rule out
ongoing CSM interaction as the primary late-time power source of
SN~2006gy. Instead, we propose that the evolution of SN~2006gy is
consistent with one of two possible scenarios: (1) A pair-instability
SN plus modest CSM interaction, where the radioactive decay luminosity
shifts to the IR because of dust formation.  (2) An IR echo, where
radiation emitted during peak luminosity heats a pre-existing dust
shell at radii near 1 light year, requiring the progenitor star to
have ejected {\it another} shell of $\sim$10 M$_{\odot}$ about 1500 yr
before the SN.

\end{abstract}


\section{Introduction and Predictions}

At late times, most supernova (SN) explosions are powered by the
radioactive decay sequence $^{56}$Ni $\rightarrow$ $^{56}$Co
$\rightarrow$ $^{56}$Fe.  Typically, less than 0.1 M$_{\odot}$ of
$^{56}$Ni is synthesized in the explosion resulting from a massive
star's core collapse, producing a light curve following the decay rate
of $^{56}$Co at late times.

Some supernovae (SNe) draw additional fuel by converting shock energy into 
light when the ejecta blast wave collides with the circumstellar medium
(CSM). The relatively slow CSM gas (a few 10$^2$ km~s$^{-1}$, compared
with $\sim$10$^4$ km~s$^{-1}$ for SN ejecta) is usually observed to be H
rich, causing bright ``narrow'' H lines in the spectrum that define the 
Type IIn subclass (Schlegel 1990; see Filippenko 1997 for a review).  
A Type IIn spectrum is
therefore a hallmark of CSM-interaction SNe. Sufficiently dense CSM
can decelerate the blast wave, draining the shock of kinetic energy
that is converted to X-rays and visual light.  In this way, SNe IIn
can be more luminous than normal SNe~II (e.g., Chevalier \& Fransson
1985; Chugai 1991, 1992).

The recent event SN~2006gy challenged this picture because it radiated
far more visual light than any other known SN (Ofek et al.\ 2007;
Smith et al.\ 2007, hereafter Paper~I).  Its soft X-ray emission
(Paper~I) and its Type IIn spectrum (Ofek et al.\ 2007; Paper~I)
showed signs of CSM interaction, but they indicated a level of
interaction that was too weak for SN~2006gy's extreme energy demands.

\subsection{CSM Interaction as the Engine}

Previously (Paper~I), we discussed both sides of this issue in detail,
noting problems with a conventional CSM-interaction scenario.  While
tell-tale signatures of {\it some level} of CSM interaction are seen
in SN 2006gy, they indicate that it is too weak by 2--3 orders of
magnitude to power the luminosity.  Thus, a traditional model of
ongoing CSM interaction, where the SN luminosity is powered by
radiation directly from the post-shock cooling zone, cannot fully
account for the energy budget of SN~2006gy during the main peak of its
light curve.  Moreover, Smith \& McCray (2007) pointed out that the
requisite CSM mass of $\ga$10 M$_{\odot}$ makes the interaction region
opaque and largely invisible.  If CSM interaction powers SN~2006gy, a
different conceptual model is needed.

Smith \& McCray (2007) argued that if 1.5$\times$10$^{51}$ erg of
ejecta shock energy were thermalized throughout an opaque envelope
with initial radius $\sim$160 AU (the pseudo photosphere of an 
unbound shell), adiabatic losses could be averted in
a manner analogous to that of most SNe~II. As the initially opaque
envelope expands and thins, trapped thermal energy can leak out.  In
this ``shell-shocked'' model, subsequent diffusion of energy can
account for the high luminosity and shape of the light curve of SN
2006gy.  A prerequisite is that the star ejected an opaque 10--20
M$_{\odot}$ shell in a luminous blue variable (LBV)-like outburst
(Smith \& Owocki 2006), fortuitously occurring in the decade before
the SN. Woosley et al.\ (2007) presented a similar CSM-interaction
model with a 25 M$_{\odot}$ pre-SN shell, wherein the precursor
outburst was triggered by the pulsational pair instability in a very
massive star, yielding model light curves approximating that of
SN~2006gy. (Note that a pulsational pair instability ejection is
different from a genuine pair-instability SN that destroys the star;
see Heger \& Woosley 2002.)

\subsection{The Pair-Instability SN Hypothesis}

The other potentially viable model to account for the high luminosity
and long duration of SN~2006gy is that it shares the same radioactive
energy source as most SNe, but that the initial mass of $^{56}$Ni is
50--100 times larger.  This would require a pair-instability SN event
(Barkat et al.\ 1967; Rakavy \& Shaviv 1968; Bond, Arnett, \& Carr
1982), where several solar masses of $^{56}$Ni are synthesized in the
explosion.  Nomoto et al.\ (2007) produced a pair-instability model
roughly matching the light curve of SN~2006gy using 15~M$_{\odot}$ of
$^{56}$Ni.

\subsection{Predictions}

The shell-shocked model and the pair-instability model make different
predictions for the late-time ($+$1 yr) luminosity, although it should
be noted that both models require the progenitor of SN~2006gy to have
been an extremely massive star, with a likely initial mass above 100
M$_{\odot}$ (Paper~I; Woosley et al.\ 2007).

While the shell-shocked model may explain the main peak of the light
curve of SN~2006gy, the energy source cannot last: as the material
continues to expand and thin out, radiation leaks away at a faster
pace and the shocked envelope suffers adiabatic losses, so the emitted
energy plummets rapidly at late times.  This hypothesis therefore
predicts that after about a year the SN would quickly fade beyond
detectability (Smith \& McCray 2007).  The model of Woosley et al.\
(2007) makes a similar prediction.  If the SN still shines brightly
after that time, there must be some continuing energy deposition.

That sustained source of energy deposition could, in principle, arise
from continued CSM interaction with additional material encountered by
the expanding blast wave after passing through the massive opaque
shell.  This would correspond to the progenitor's normal wind before
the precursor shell ejection.  This type of ongoing CSM interaction is
the late-time engine for most SNe~IIn that remain luminous.  In all
such cases, however, the lasting luminosity is accompanied by very
strong, relatively broad H$\alpha$ emission. This is {\it not} the
case for SN~2006gy, as we will demonstrate in this paper, making the
ongoing CSM interaction hypothesis improbable.

The pair-instability SN hypothesis predicts that the luminosity will
decline slowly, following the $^{56}$Co decay rate or slightly faster,
so it should remain luminous.  An important point is that the decay
rate can be faster than the intrinsic $^{56}$Co decay rate if the
material becomes optically thin, and this effect may be important in
allowing a self-consistent pair-instability explanation for SN~2006gy.

The different expectations for radioactive decay and the shell-shocked
model would seem to provide a straightforward test: if SN~2006gy faded
rapidly, it was not a pair-instability SN.  By day 230 after
explosion, SN~2006gy appeared to be fading slower than expected in the
shell-shocked model, but thereafter it became lost in the Sun's glare.
Here we set out to test the aforementioned predictions by attempting
to detect SN~2006gy at late times after it re-emerged from behind the
Sun.  As we shall see, the surprising results do not fully solve the
mystery, and they introduce a new twist to the story.

\begin{figure}
\epsscale{1.0}
\plotone{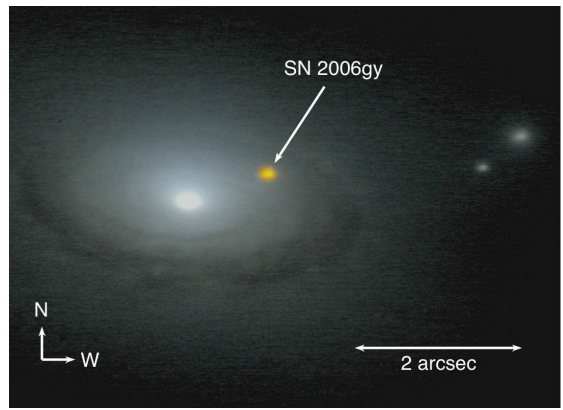}
\caption{ False-color IR image of the core of NGC\,1260 and SN\, 2006gy,
generated from Keck LGS/NIRC2 $H$-band and $K^\prime$-band observations on
2007 Dec. 1.  SN 2006gy is red and very luminous in the near-IR, 
more than a year after explosion.}
\end{figure}

\section{Observations}

Once SN~2006gy became observable again, we began a monitoring campaign
of unfiltered optical photometry with the Katzman
Automatic Imaging Telescope (KAIT; Filippenko et al.\ 2001), $R$-band
photometry with the low-resolution imaging spectrometer (LRIS; Oke et
al. 1995) on the Keck~I 10-m telescope, and $K\arcmin$ and $H$
photometry with the Near-Infrared Camera 2 (NIRC2) and the laser guide
star (LGS) adaptive optics (AO) system (Wizinowich et al.\ 2006) on
Keck~II.  Two optical spectra were also obtained with LRIS on Keck~I.

The unfiltered KAIT photometry resumed on 2007 Aug. 17 (day 362; UT
dates are used throughout this paper), but we could only establish
limits due to its position in the bright inner regions of its host
galaxy.  Using fake-star injection tests on images with host galaxy
light subtracted, we estimate that SN~2006gy had an unfiltered
apparent magnitude fainter than about 19.5 on days 362 and 394, or an
absolute magnitude fainter than $-$16.5 with the same parameters as in
Paper~I.  Similarly, we estimate an $R$-band magnitude fainter than
20.0 from the Keck image obtained on 2007 Oct. 16 (day 422).

Using NIRC2 LGS AO, we obtained $K\arcmin$ photometry on 2007 Sep.\ 29
in wide-field mode, and $K\arcmin$ and $H$-band photometry on 2007
Dec.\ 1 in narrow-field mode.  SN~2006gy was clearly detected in all
bands at both epochs (Fig.\ 1).  Based on a comparison with the only
two well-isolated 2MASS stars in common within our small field, we
measured $K\arcmin = 15.1 \pm 0.1$ mag on day 405, and $K\arcmin =
15.4 \pm 0.1$ mag on day 468.  The $H$-band photometry could not be
calibrated owing to the even smaller field of view for the Dec.\ 1
observation, in which there were no good 2MASS comparison stars. (The
$K\arcmin$ photometry on Dec.\ 1 could be reliably flux calibrated,
however, because of field stars in common with the larger
field-of-view $K\arcmin$ image obtained on Sep.\ 29.)\footnote{Figure
1 also shows clear evidence for a dark dust lane at a projected radius
of $\sim$2\arcsec\ or 700 pc from the nucleus of NGC~1260.  This
confirms structures seen by Ofek et al.\ (2007), and reinforces the
possibility that the host galaxy has active massive star formation.}

Given the questions concerning CSM interaction as a potential energy
source at late times, we also obtained a deep optical spectrum to
search for any lingering intermediate-width H$\alpha$ emission coming
from the shocked CSM gas.  Our LRIS spectra were obtained on 2007
Aug.\ 19 and Oct.\ 16, and were reduced using standard techniques
(e.g., Foley et al.\ 2003).

\begin{figure}
\epsscale{1.0}
\plotone{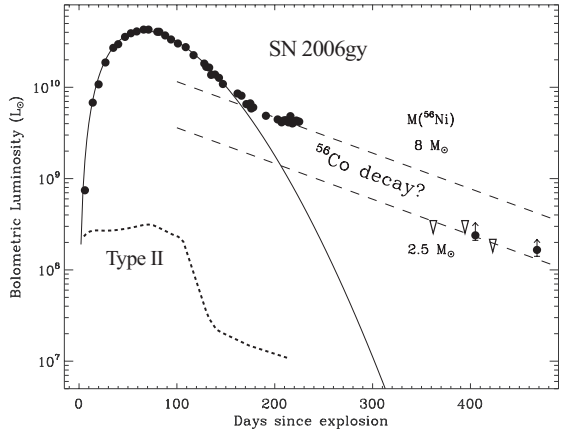}
\caption{Bolometric luminosity light curve of SN~2006gy.  The data
points up to day 230 are from the absolute $R$ magnitudes in Paper~I,
with no bolometric correction.  The solid curve is the
``shell-shocked'' photon diffusion model of Smith \& McCray (2007).
The two late-time data points are lower limits to the luminosity
derived from our observed $K\arcmin$-band magnitudes, assuming that
the SED peaks in the $K\arcmin$ band.  Dashed lines show the expected
energy deposition rate for representative masses of $^{56}$Ni,
assuming 100\% efficiency of the ejecta in absorbing radioactive decay
luminosity (this may be optimistic, and so the $^{56}$Ni mass
indicated by the observed luminosity may be higher).  Since the
late-time IR luminosities are lower limits, we emphasize that 2.5
$M_{\odot}$ is only a lower bound for the $^{56}$Ni mass in the
radioactive decay hypothesis. Similarly, while unfilled triangles
denote KAIT and Keck/LRIS $R$-band upper limits, they are not really
limits to the bolometric luminosity because the SED does not peak in
the $R$ band.}
\end{figure}

\section{Results and Discussion}

\subsection{Minimum Luminosity from $K^\prime$ Photometry}

Although we do not have sufficient color information for an accurate
assessment of the bolometric luminosity, we can derive a {\it minimum}
luminosity of SN 2006gy at late times assuming that a blackbody-like
energy distribution peaks in the $K\arcmin$ band (no bolometric
correction to the absolute $K\arcmin$ magnitude).  This is plausible,
given its red $H-K^\prime$ color compared to the host galaxy (Fig.~1).
The bolometric luminosity could be significantly higher than what we
quote below, which is why we plot lower limits in Figure 2.  For
example, if the spectral energy distribution (SED) is not a
single-temperature blackbody, then the luminosity will be higher.  If
it peaks at longer IR wavelengths near 5~$\micron$, as we may expect
(see below), then the bolometric luminosity could be 5--10 times
larger.  Upper limits corresponding to $R$-band nondetections are also
plotted in Figure 2; these do not provide stringent constraints, but
they also do not conflict with luminosities derived from the
$K\arcmin$ band. They assume that the SED peaks in the $R$ band, which
is unlikely; thus, the true upper limits are probably substantially
higher than shown.

Adopting the same 73.1 Mpc distance and $A_R=1.68$ mag ($A_K=0.24$
mag) extinction as in Paper~I, we find minimum values for the
bolometric luminosity of $\sim2.4\times10^{8}$ L$_{\odot}$ on day 405
and $\sim1.7\times10^{8}$ L$_{\odot}$ on day 468 from the $K\arcmin$
measurements.  If the SED does not peak at $\sim$2.3 $\micron$, then
the true luminosity is higher.

Figure 2 shows lower limits to the late-time luminosity, plotted along
with the light curve and photon-diffusion model from Paper~I and Smith
\& McCray (2007), respectively.  The measured late-time bolometric
luminosity is much higher than the predicted decay from photon
diffusion in the ``shell-shocked'' model of Smith \& McCray (2007), as
well as that of Woosley et al.\ (2007).  The additional luminosity
source could be either (1) continued CSM interaction as the shock runs
into an extended, dense CSM created by a progenitor with $\dot{M}
\approx 10^{-2}$ M$_{\odot}$ yr$^{-1}$ (see \S 3.2), (2) radioactive
decay from $\ga$2.5 M$_{\odot}$ of $^{56}$Ni (\S 3.4), or (3) an IR
echo, as light from the time of peak luminosity is now heating
dust in {\it another} massive shell at a radius of $\sim$1 light year
from the SN, ejected by the progenitor star $\sim$1500 yr earlier.

\begin{figure}
\epsscale{1.1}
\plotone{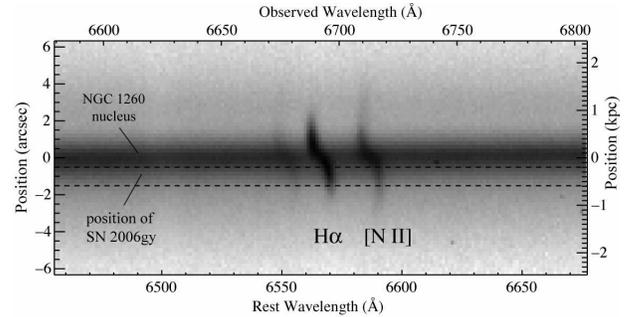}
\caption{The observed spectrum at the position of SN~2006gy on day 364
  after explosion; the SN position is halfway between the two dashed
  lines.  Continuum from the nucleus of NGC~1260 is seen clearly, as
  is the narrow H$\alpha$ and [N~{\sc ii}] emission from H~{\sc ii}
  regions that follow the galaxy's rotation curve.  We detect no
  broad or intermediate-width H$\alpha$ emission from SN~2006gy.}
\end{figure}

\subsection{CSM Interaction and the Lack of H$\alpha$ Emission}

One hypothetical source for a slowly declining luminosity, especially
for SNe~IIn, is continued CSM interaction as the blast wave encounters
{\it additional} dense material outside of the LBV-like ejecta shell
required in the ``shell-shocked'' models of Smith \& McCray (2007) and
Woosley et al.\ (2007).  A convenient expression for the progenitor's
minimum mass-loss rate needed to produce an observed luminosity $L_9 =
L_{SN}/(10^9~{\rm L}_{\odot})$ through CSM interaction, with an optimistic
100\% efficiency, is given by

\begin{displaymath}
\dot{M} = 0.04 \ L_9 \big{(} \frac{v_{w}}{200} \big{)} \big{(} 
\frac{v_{SN}}{4000} \big{)}^{-3} ~{\rm M}_{\odot}~{\rm yr}^{-1},
\end{displaymath}

\noindent where $v_{w}$ and $v_{SN}$ are the progenitor's wind speed
and the SN blast-wave speed (respectively) in km s$^{-1}$.  For the
main peak of SN~2006gy, the observed luminosity of $L_9 = 50$ at 70~d
after explosion requires a mass-loss rate for the progenitor of at
least 2 M$_{\odot}$ yr$^{-1}$ for 5--10 yr before explosion.
Similarly, a late-time luminosity of $L_9 = 0.2$--0.3 at $\sim$400~d
requires at least $\dot{M} \approx 10^{-2}$ M$_{\odot}$ yr$^{-1}$ for
$\ga$30 yr before the SN, if that luminosity arises from CSM
interaction.

There are two interesting aspects of this derived mass-loss rate.
First, it matches the value we found in Paper~I from the luminosity of
the narrow H$\alpha$ component, tracing the (at that time) pre-shock
ionized wind.  This would seem suspicious if it were merely a
coincidence.  Second, it is still {\it extremely high} compared to the
mass-loss rates measured for steady stellar winds of normal supergiant
stars.  A mass-loss rate of $10^{-2}$ M$_{\odot}$ yr$^{-1}$ --- while
much lower than the value of $\sim$1 M$_{\odot}$ yr$^{-1}$ for
$\eta$~Carinae during its 1840s eruption (Smith et al.\ 2003) and for
SN~2006gy in the decade preceding the SN (Paper~I) --- is still about
100 times more than can be produced by a line-driven wind, and is
representative of smaller LBV eruptions like that of P~Cygni in 1600
A.D.\ (Smith \& Hartigan 2006).  Thus, CSM interaction requires two
extraordinary phenomena: (1) an extreme giant LBV-like eruption such
as $\eta$~Car in the 5--10 yr before explosion, plus (2) a sustained
but less extreme eruption for several decades before that.

Some SNe of Types IIn and II-L decline very slowly, having relatively
high luminosities even at late times, $\ga$1 yr after discovery.  Two
well-studied examples are SN~1988Z and SN~1979C (see Filippenko 1991,
1997; Turatto et al.\ 1993; Branch et al.\ 1981), with CSM interaction
proposed as the cause of their long duration.  Both were more luminous
than a normal SN~II, although still over 10 times less luminous than
SN~2006gy.  SN~1988Z also had strong radio and X-ray emission (Van Dyk
et al.\ 1993; Fabian \& Terlevich 1996). In both objects, {\it broad
or intermediate-width H$\alpha$ emission continued to be very strong
and easily detected for many years, as long as the SNe remained
luminous}.  In SN~1988Z, the H$\alpha$ luminosity peaked $\sim$1~yr
after discovery and declined slowly thereafter (Turatto et al.\ 1993).
In fact, bright H$\alpha$ dominating the visual spectrum is seen in
all other SNe~II where CSM interaction powers the late-time luminosity
(e.g., SN~1980K, SN~1986J, SN~1987F, SN~1993J; Leibundgut et al.\
1991; Chugai 1990; Matheson et al.\ 2000).  Recently, Milisavljevic et
al.\ (2008) show that post-shock H$\alpha$ emission from SN~1986J is
{\it still} detectable.  Similarly, the SN~IIn 2006tf, which was
contemporaneous with SN~2006gy and almost as luminous, also shows
easily detected broad H$\alpha$ emission in its late-time spectrum
obtained 1.5 yr after explosion (Smith et al.\ 2008b).  In general,
the presence of strong, relatively broad H$\alpha$ emission is a 
robust prediction for late-time CSM interaction (Chugai 1990, 1991).

Figure 3 shows the red spectrum at the position of SN~2006gy.  We
detect continuum emission from the nucleus of NGC~1260, and narrow
H$\alpha$ and [N~{\sc ii}] from H~{\sc ii} regions. However, {\it
there is no hint of broad or intermediate-width H$\alpha$ emission at
the position of SN~2006gy}.  The failure to detect the $R$-band
continuum luminosity could conceivably be blamed on spatial
resolution, but not the lack of relatively broad H$\alpha$ emission.
In order to go undetected in our LRIS spectrum in Figure 3, the
broad/intermediate H$\alpha$ flux must be less than the limiting
background continuum level, which is roughly 2$\times$10$^{-17}$ erg
s$^{-1}$ cm$^{-2}$ \AA$^{-1}$ at 20 mag.  If the H$\alpha$ component
has a Doppler width of $\sim$1000 km s$^{-1}$, this implies that the
upper limit to the late-time H$\alpha$ luminosity of SN~2006gy is
$\sim$10$^{39}$ erg s$^{-1}$.  To put this in perspective, the
integrated H$\alpha$ luminosity of SN~2006gy must be about {\it 400
times weaker} than those of SNe~2006tf and 1988Z at a similar epoch
(Turatto et al.\ 1993; Smith et al.\ 2008b), even though the late-time
bolometric luminosity of SN~2006gy was comparable to that of
SN~2006tf, and about 10 times stronger than that of SN~1988Z.

Moreover, if continued CSM interaction were the power source,
SN~2006gy should be extremely luminous in X-ray and (eventually) radio
emission, analogous to SN~1988Z.  However, X-ray observations of
SN~2006gy obtained with {\it Chandra} on 2007 Dec.\ 15 were unable to
detect SN~2006gy, with a preliminary analysis yielding an estimated
upper limit to the 0.5--2 keV X-ray luminosity of 1.4$\times$10$^{39}$
erg s$^{-1}$ (Pooley et al.\ 2008, in prep.). This X-ray luminosity is
about 100 times less than that of SN~1988Z (Fabian \& Terlevich 1996).
SN~2006gy has faded compared to our previous X-ray observation
(Paper~I), and even the mass-loss rate of $\sim$10$^{-4}$ M$_{\odot}$
yr$^{-1}$ implied by those earlier observations would have been 100
times too low to account for its late-time bolometric luminosity.
Furthermore, no radio detection of SN~2006gy has been reported yet,
and available upper limits on days 263, 265, 308, 434, and 538
(Bietenholz \& Bartel 2007; Argo et al.\ 2007; Bietenholz 2008,
priv. comm.) make it significantly less radio luminous than a typical
SN~IIn.  Altogether, we find the case for ongoing CSM interaction as
the dominant late-time luminosity engine of SN~2006gy to be quite
weak.

\subsection{Can Extinction from New Dust Resurrect the CSM-Interaction
Hypothesis?}

The bulk of the bolometric flux of SN 2006gy has shifted into the IR,
and we have detected no visual-wavelength emission from the SN.  One
might ask if dust formation could obscure the optical continuum and
H$\alpha$ emission.  Given the large amount of CSM mass inferred for
this object (Paper~I; Smith \& McCray 2007; Woosley et al.\ 2007),
dust formation might indeed have occurred in the dense post-shock
shell, in a manner analogous to that of SN~2006jc (Smith et al.\
2008a).  The dust mass required to emit the observed $K\arcmin$-band
luminosity (assuming $T=1300$~K, so that the energy distribution peaks
in the $K\arcmin$ band) is $\sim 10^{-3}$ M$_{\odot}$ (adopting grain
opacities in Draine 2003).  At $R \approx 10^3$ AU, this mass of dust
could provide $\tau \approx 4$ at red wavelengths, insufficient to
hide an H$\alpha$ luminosity equivalent to that of SN~1988Z.

Even so, some fine tuning would be required for newly formed dust to
obscure the H$\alpha$ emission and to re-emit the CSM-interaction
luminosity.  This is because the ongoing CSM interaction needed to
power the late-time luminosity must occur {\it outside} the densest
swept-up shell where the dust is most likely to form.  The ongoing CSM
interaction would occur at large radii as the shock plows into the gas
lost before the LBV-like precursor outburst, as discussed above.
Thus, there is no natural way for dust formed in the massive shocked
shell (now well inside the blast wave) to block our view of that {\it
external} CSM-interaction, and dust would be very inefficient at
absorbing and re-emitting that shock luminosity.  Furthermore, dust in
any geometry could not explain the lack of radio emission noted
earlier.

On the other hand, obscuration from dust that may have formed poses no
difficulty for the pair-instability SN hypothesis.  If sufficient dust
formed in the dense shell or massive stellar envelope, it could
conceivably block our view of the dominant {\it internal} energy
source, which is radioactive decay from the central ejecta in that
hypothesis.  It would act as a calorimeter of that radioactivity,
absorbing the luminosity and re-emitting it at IR wavelengths, as
observed.  If the radius of the shell is $\sim$10$^3$ AU ($v_{SN}
\approx 4000$ km s$^{-1}$ for 1 yr), then for the current luminosity
of $\ga$2$\times$10$^8$ L$_{\odot}$, the equilibrium grain temperature
is $\ga$1000 K, giving a self-consistent explanation for the 2
$\micron$ flux.

\subsection{A Large $^{56}$Ni Mass and the Pair-Instability SN Hypothesis}

While the ongoing CSM-interaction hypothesis seems to have
debilitating obstacles, the late-time luminosity and decline rate with
weak H$\alpha$ emission seem entirely consistent with a luminosity
powered by radioactive decay from a large mass of $^{56}$Ni.  If dust
formed, as the IR data imply regardless of the luminosity source, then
the shift to IR wavelengths can also be
explained.\footnote{Alternatively, the IR shift may have been caused
by hyperfine transitions in the cooled ejecta rather than continuum
emission, analogous to the ``IR catastrophe'' predicted for SNe~Ia
(Axelrod 1980). IR spectra could be used to test this hypothesis.}

The minimum luminosity derived from late-time IR magnitudes suggests
that SN~2006gy is fading slowly.  This decay rate is roughly
consistent with that of $^{56}$Co (Fig. 2), as long as there is no
strong color evolution between our two $K\arcmin$ observations.

According to Figure 2, the {\it minimum} necessary $^{56}$Ni mass is
2.5 M$_{\odot}$.  This seems insufficient to account for the peak
luminosity of SN~2006gy, which requires a higher $^{56}$Ni mass of
10--20 M$_{\odot}$ (Paper~I; Nomoto et al.\ 2007).  However, the
actual decline rate can be somewhat faster than the intrinsic
$^{56}$Co decay rate if the ejecta become partly optically thin,
underscoring again that 2.5~M$_{\odot}$ is a lower limit.  In fact, in
the pair-instability SN model that Nomoto et al.\ (2007) presented for
SN~2006gy, representing the explosion of a star with an initial mass of
166~M$_{\odot}$, the late-time luminosity decline rate was faster than
that of $^{56}$Co.  With a $^{56}$Ni mass of 15~M$_{\odot}$, the model
of Nomoto et al.\ was able to give a satisfactory fit to the main
light-curve peak, and it {\it predicted} a late-time luminosity that
is within $\sim$30\% of the observed value we present here.  Thus, to
the extent that such models are a fair representation of a 
pair-instability SN in the modern Universe (the model of Nomoto et al. 
did include significant mass loss from the progenitor, appropriate for
the near-solar metallicity of SN~2006gy; Ofek et al.\ 2007; Paper I), it
seems that the pair-instability SN hypothesis for SN~2006gy remains
entirely valid, as long as dust can form in order to shift the energy
distribution into the IR.

Although the estimate of 2.5 M$_{\odot}$ is a lower limit, it could
also be the case that the main peak of SN~2006gy is powered largely by
optically thick CSM interaction (Smith \& McCray 2007; see also Ofek
et al.\ 2007; Paper~I; Woosley et al.\ 2007), while the late-time
luminosity may be powered by radioactive decay of $^{56}$Co.  These
two possibilities are not mutually exclusive.  The $^{56}$Ni mass of
2.5 M$_{\odot}$ is, in some sense, more reasonable than 10--20
M$_{\odot}$ of $^{56}$Ni, because of the relaxed constraints on the
initial stellar mass (Heger \& Woosley 2002).  This is roughly 4 times
more than can be produced in energetic core-collapse SNe, for which
Woosley \& Weaver (1995) find maximum $^{56}$Ni yields of $\la$0.7
M$_{\odot}$.

\begin{figure}[!ht]
\epsscale{1.0}
\plotone{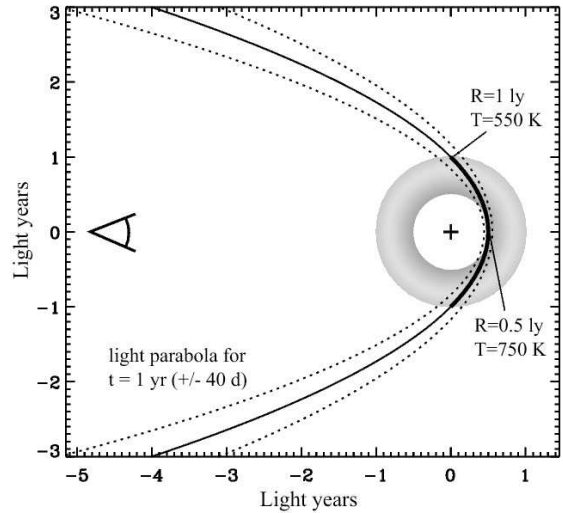}
\caption{The possible geometry of the IR echo of SN~2006gy.  The solid
  curve is the light parabola cone for a time $\sim$1 yr after the
  luminosity peak, and the dotted curves are the corresponding
  parabolas for 40 days before and after that.  The intersection of
  these curves and the circular shell represents the locus of points
  in the CSM that we see illuminated by the object's high-luminosity
  phase.  The closest material illuminated is at a radius of 0.5 ly
  from the star.  Material much farther away can be reached by the
  light cone, but in order to be hot enough to emit thermal radiation
  at $\sim$2~$\micron$, it cannot be much farther than 1 ly from the
  star.  Thus, the thick part of the curve represents the probable
  location of the dust responsible for the IR echo of SN~2006gy, and
  the shaded thick circle represents the rest of a shell that is not
  illuminated by the SN but which may cause extinction along our
  sight-line.}
\end{figure}

\subsection{An IR Echo From Another Massive LBV Shell}

Some SNe~IIn show strong near-IR emission $\ga$1 yr after explosion
(e.g., Gerardy et al.\ 2002), usually interpreted in the framework of
an IR echo, where UV and visual radiation from the SN at peak light
has reached a dust shell at large radii.  Thus, the dust is heated
and reradiates the absorbed energy in the IR (Dwek 1983).  A similar
model could potentially explain the late-time IR luminosity of
SN~2006gy, since its substantial reddening (Ofek et al.\ 2007;
Paper~I) implies a significant amount of CS dust along our
line of sight.

At about 1 yr after peak luminosity, we may see IR reradiation from
heated dust located at the intersection where a light parabola
corresponding to $t = 1$ yr after the SN flash passes through a
massive dusty shell at radii of 0.5--1 ly from the SN (Fig.\ 4). For a
peak luminosity of 5$\times$10$^{10}$ L$_{\odot}$ (Fig.\ 2), the
equilibrium grain temperatures at those radii should be roughly
550--750 K.  There may be additional dust at even larger radii along
the light cone, but it will be even cooler and will not contribute
substantially to the $K\arcmin$-band flux (Fig.\ 4).  Thus, if this
interpretation is correct, the SED should peak at wavelengths of 4--5
$\micron$ or longer.

The IR echo model has two important consequences: (1) the $JHK$ and
mid-IR photometry should show an SED rising toward longer wavelengths
plus a temperature that drops with time, and (2) the true dust
luminosity is probably 5--10 times higher than our estimate in Figure
2, or at least 10$^9$ L$_{\odot}$, because the $K$ band is on the Wien
tail of a blackbody that peaks at 5 $\micron$.  With standard
assumptions about dust grains at those radii and temperatures, the
dust mass needed to account for that IR luminosity is at least
0.05--0.1 M$_{\odot}$.  The true dust mass in the shell must be vastly
more, because this estimate only corresponds to the heated dust near
the light cone (Fig.\ 4), which represents a small fraction of the
total dust-filled volume.  With a normal gas-to-dust ratio, this
requires the existence of a pre-existing shell of more than 5--10
M$_{\odot}$.  At a radius of 1~ly, with a likely expansion speed of
200 km s$^{-1}$ (Paper~I), this shell would have been ejected by the
progenitor star $\sim$1,500 yr ago.\footnote{The volume interior to
this dust shell, but exterior to the current forward shock radius, is
presumably filled with a steady low-density wind blown between the two
massive shell ejections.  We have no reliable constraints on this
material, however, because it would have been illuminated when
SN~2006gy was unobservable.}


If this alternative IR echo hypothesis is correct, it greatly
strengthens the comparison that we drew in Paper~I between the
progenitor of SN~2006gy and the massive star $\eta$
Carinae. Specifically, in Paper~I we proposed this similarity based
mainly on the requirement from the observed light-curve peak that it
ejected 10--20 M$_{\odot}$ of H-rich gas in the decade immediately
before the SN.  Now, with an IR echo, we find that the progenitor of
SN~2006gy may have suffered violent pre-SN mass-loss episodes at least
{\it twice}, with a previous LBV-like eruption occurring about 1500~yr
ago.  We did not predict this behavior for SN~2006gy, but it is
something that is already well-established for $\eta$ Car: in addition
to its 10--20 M$_{\odot}$ nebula (Smith et al.\ 2003) that was ejected
in the 19th-century eruption, $\eta$ Car has multiple shells at larger
radii {\it of the same size inferred here}, indicating previous
massive-shell ejections roughly 1000--2000 yr ago (Walborn 1976;
Walborn et al.\ 1978; Smith \& Morse 2004; Smith et al.\ 2005).

If this massive, older LBV-like shell exists around SN~2006gy, the
observed bolometric light curve would not necessarily disagree with
the original opaque shell-shocked model of Smith \& McCray (2007),
because now there is a different explanation for the late-time
luminosity. This may pose some difficulty for the pulsational pair
instability model of Woosley et al.\ (2007) as the explanation for the
pre-SN mass ejection, however, which predicts a giant mass ejection in
the decades before core collapse, as in SN~2006gy, {\it but not
$\sim$1000 yr before that}.  With two successive ejected shells of
10--20 M$_{\odot}$ each, alternative pre-SN outburst mechanisms that
have been proposed involving Ne flashes from degenerate cores of
$\sim$11 M$_{\odot}$ stars (Chugai \& Danziger 2003; Weaver \& Woosley
1979) are obviously ruled out in this particular case. (It should be
noted that such flashes may have been a numerical artifact anyway,
since they do not occur in more recent models; Woosley et al.\ 2002.)

Instead, the IR echo hypothesis for the late-time luminosity of
SN~2006gy would imply that the precursor ejections of SN~2006gy were
more akin to sporadic, giant LBV eruptions (Smith \& Owocki 2006), one
of which happened to occur (or was induced to occur) just before the
explosion.  If that link is valid, the implications for $\eta$ Car are
provocative, and renewed efforts to diagnose the causes of such
eruptions should be made a high priority for stellar evolution theory.
Finally, if the IR echo model can fully account for the late-time IR
luminosity, and if better limits can be placed on the optical
luminosity, then this weakens the motivation for a large mass of
$^{56}$Ni and the pair-instability hypothesis.

\section{Summary}

Interpretation of the late-time properties of SN 2006gy is admittedly
complicated, adding to an already complex story from its early-time
behavior.  This is not so unusual in astronomy, when new observations
of extreme objects push the limits of known physical parameters.

None of the three late-time luminosity sources that we have discussed
match predictions of models for the early-time luminosity, and each
has the unappealing property that it requires us to introduce some new
aspect to the problem.  The late-time CSM-interaction hypothesis
requires us to invoke additional mass loss and asks us to ignore the
extremely weak H$\alpha$, X-ray, and radio emission that are expected
from the observed luminosity; the radioactive decay hypothesis
requires new dust formation to shift the SED into the IR; and the
IR-echo hypothesis requires that the progenitor star had ejected {\it
another} shell of $\gtrsim$10 M$_{\odot}$ about 1500 yr before the SN,
in addition to the one required 5--10 yr before for the main light
curve peak.  In light of these complications, it is useful to
highlight some reasonably definite results from our analysis of the
late-time properties of SN 2006gy.

First, we can rule out a model invoking traditional CSM interaction,
where direct radiative cooling from ongoing shock interaction powers
either the early-time or the late-time luminosity of SN~2006gy.  This
is due to the fact that post-shock H$\alpha$ emission, X-rays, and
radio emission are all orders of magnitude weaker than expected in
such a model. In this respect, SN~2006gy is unique compared with other
SNe~IIn having similar longevity in their luminosity.  We would not be
surprised if there is {\it some} level of ongoing CSM interaction, of
course, but our limits show that any continued CSM interaction is too
weak to power the observed late-time IR luminosity.  Because of the
geometry, it is also difficult to explain the shift to IR wavelengths
in a standard CSM-interaction scenario.

Second, on a more encouraging note, we can narrow down the possible
interpretations of SN~2006gy to two fairly well-defined alternative
models:

(1) So far, all observations seem consistent with the hypothesis that
SN~2006gy was a genuine pair-instability SN, as long as we permit dust
formation that shifts the late-time SED to longer IR wavelengths.  For
example, the model by Nomoto et al.\ (2007) for a star of initial mass
166 M$_{\odot}$ that explodes as a pair-instability SN producing 15
M$_{\odot}$ of $^{56}$Ni can explain both the main peak of the light
curve {\it and} adequately accounts for the late-time luminosity we
present here.  The shift to IR wavelengths may be plausible too, since
dust formation is {\it predicted} in some models for pair-instability
SNe (Schneider et al.\ 2004).  The level of CSM interaction indicated
by the observed Type IIn spectrum, the H$\alpha$ luminosity, and the
observed X-ray emission require that the progenitor star had a modest
mass-loss rate comparable to those of extremely luminous blue
supergiants (Paper~I), but this is not necessarily surprising since
SN~2006gy occurred in an environment that was not metal poor (Paper~I;
Ofek et al.\ 2007).  In fact, the model by Nomoto et al.\ (2007) did
invoke significant pre-SN mass loss in order to explain the
light-curve shape.

(2) The second plausible scenario is that the progenitor of SN~2006gy
    ejected a series of extremely massive H-rich shells (at least
    twice) before the explosion.  Both of these shells had likely
    masses of 10--20 M$_{\odot}$ or more. One was ejected within a
    decade before the SN, while the other was ejected $\sim$1,500 yr
    earlier if it had the same expansion speed of roughly 200 km
    s$^{-1}$.  The 10--20 M$_{\odot}$ shell ejected in the decade
    preceding the SN and residing at a radius of
    $\sim$2$\times$10$^{15}$ cm is needed to explain the luminosity
    and light-curve shape at early times via an opaque diffusion model
    (Smith \& McCray 2007; Woosley et al.\ 2007), whereas we infer
    that the shell of comparable mass ejected 1,500 yr earlier is
    needed to account for the late-time IR luminosity in the context
    of a thermal-IR echo from an extended dust shell at a radius of
    0.5--1 ly.

If the first hypothesis is correct and SN~2006gy was indeed a
pair-instability SN, then important implications follow.  These SNe
have thus far been discussed mainly in the context of very massive
stars in the early Universe at very low metallicity.  However, key
assumptions about metallicity-dependent mass loss are not necessarily
reliable for very massive stars, as pointed out by Smith \& Owocki
(2006).  SN~2006gy would also imply that the nominal requirements of
low metallicity for pair-instability SNe are relaxed, and that such
explosions could occur at all epochs given a sufficiently massive star
(see Paper~I for additional discussion).

On the other hand, it could be argued that of the two hypotheses noted
above, both of which seem plausible, the IR echo from a massive
extended shell is arguably the more compelling, because its
requirements have some precedent in the observed mass-loss history of
stars like $\eta$~Carinae (e.g., Smith et al.\ 2003).  We note that
Barlow et al.\ (2005) also invoked a similar distant dust shell of
$\sim$10~M$_{\odot}$ in order to account for the IR echo of SN~2002hh.
Since the progenitor of SN~2006gy may have suffered {\it two} violent
mass ejections of 10 M$_{\odot}$ or more separated by 1,500 yr, it is
now unclear whether the episode in the decade immediately preceding
SN~2006gy was tied directly to the final burning phases leading up to
the SN explosion, or instead, if it may have been a coincidence that
they occurred nearly simultaneously.  In any case, these multiple
violent mass ejections represent a fundamental challenge to our
understanding of the late evolution and deaths of massive stars, and
deserve a high priority in work on stellar evolution theory.

\acknowledgments \footnotesize

We thank an anonymous referee and M.\ Shull for helpful comments.
This study is based largely on data obtained at the W. M.\ Keck
Observatory, operated as a scientific partnership among the California
Institute of Technology, the University of California, and NASA.  The
Observatory was made possible by the generous financial support of the
W. M.\ Keck Foundation. We thank the Keck staff for their assistance.
KAIT and its ongoing research were made possible by generous donations
from Sun Microsystems, Inc., the Hewlett-Packard Company, AutoScope
Corporation, Lick Observatory, the National Science Foundation (NSF),
the University of California, the Sylvia and Jim Katzman Foundation,
and the TABASGO Foundation. A.V.F.'s supernova group at UC Berkeley is
partially supported by NSF grant AST--0607485 and NASA/Chandra grant
DD7-8041X.  This work was also partially supported by DOE-SciDAC grant
DE-FC02-06ER41453.  J.S.B.\ was partially supported by a grant from
the Hellman Faculty Fund.  P.J.M.\ is supported by a research
fellowship from the TABASGO Foundation.


\end{document}